\documentstyle[preprint,aps,psfig,floats]{revtex} 
\begin{document}
\tightenlines
\hsize\textwidth\columnwidth\hsize\csname@twocolumnfalse\endcsname

\draft

\title{Band-Structure Effects in the Spin Relaxation of Conduction Electrons}
\author{J. Fabian and S. Das Sarma}
\address{Department of Physics, University of Maryland at College Park,
College Park, MD 20742-4111}
\maketitle

\vspace{1cm}

\begin{abstract}
Spin relaxation of conduction electrons in metals is significantly
influenced by the Fermi surface topology. Electrons near 
Brillouin zone boundaries, special symmetry points, or accidental
degeneracy lines have spin flip rates much higher than an average
electron. A realistic calculation and analytical estimates show that
these regions dominate the spin relaxation, explaining why polyvalent
metals have much higher spin relaxation rates (up to three orders of
magnitude) than similar monovalent metals. This suggests that spin
relaxation in metals can be tailored by band-structure modifications
like doping, alloying, reducing the dimensionality, etc.
\end{abstract}
\vspace{2em}
\newpage

The ability of normal metals to carry a spin-polarized current
has led to the development of spin transistor\cite{johnson93a},
a  ferromagnet-normal metal-ferromagnet
sandwich device that can switch current
depending on the relative orientation of the magnets.
Spin polarized currents
also flow in the metallic layers of giant magnetoresistance
heterostructures\cite{prinz95},
substances promising for magnetic signal processing. A major
limitation of the quality of these devices is imposed by the time
(the so called spin relaxation time $T_1$\cite{pines55}) an
unbalanced spin population in metals persists; the metallic layers
used in these devices cannot be thicker than the spin diffusion
length which is proportional to $\sqrt{T_1}$. Finding ways of
enhancing $T_1$ is technologically important also from the
perspective of quantum computing that represents bits
by electronic spins\cite{divincenzo95}. One recent successful
attempt to increase $T_1$ in semiconductors by doping is described
in Ref. \cite{kikkawa97}.

We recently reported\cite{prl} on a theoretical study of the spin
relaxation in polyvalent metals where we showed how the band
structure affects $T_1$. The physical picture is the one of a random
walk on the Fermi surface: the weakness of the spin-flipping
interactions in a typical metal ensures that an
electron changes its momentum many times (typically a ten
thousand) before its spin flips. At some points (we call them {\it
spin hot spots}), however, the spin-flip probability 
is much enhanced.  Near a Brillouin zone boundary
(Bragg plane), for
example, this probability can increase a hundred times. If the
electron jumps in or out of a special symmetry or accidental
degeneracy point, the probability may be even close to one, that is,
the spin-flipping and spin-conserving jumps may be almost equally
frequent!  Although an electron jumping into a spin hot spot is a
rare event, it may nevertheless dominate the spin relaxation.

The above picture solves a longstanding experimental puzzle
formulated first by Monod and Beuneu\cite{monod79}: Why spin in some
metals decay unexpectedly fast? Experiments show that alkali and
noble metals have spin relaxation times consistent with the predictions
of the theory of Elliott\cite{elliott54} and Yafet\cite{yafet63},
when the spin-flip rates are calculated from the atomic state
parameters.  Metals Al, Pd, Mg, and Be, on the other hand, have spin
relaxation times much smaller (up to three orders of magnitude for Mg
and Be) than estimated. This discrepancy is disturbing since it
shows that otherwise similar metals may have very different spin
relaxation times. Silsbee and Beuneu\cite{silsbee83} were first to
notice that accidental degeneracy points can significantly increase
the spin relaxation of aluminum. We substantiated this idea
by a rigorous calculation, and extended it to include all 
band-structure degeneracies. 
The experimental puzzle is then resolved by recognizing
the fact that the two groups of metals have different valency.
Alkali and noble metals are monovalent--their
Fermi surface lies completely within the first Brillouin zone.
The spin-flip rates are more or
less uniform with the values close to the ones derived from atomic
physics (there are no spin hot spots).  On the other hand
polyvalent Al, Pd, Mg, and Be have numerous Fermi surface anomalies
were spin hot spots can be formed. Aluminum, for example, have
spin relaxation rates determined primarily by the Fermi surface
regions near the Brillouin zone boundaries and around the accidental
degeneracy points\cite{prl,silsbee83}.
In beryllium, the Fermi surface cuts through a degeneracy plane!

 Since no polyvalent metals other than
Al, Pd, Mg, and Be have been measured for $T_1$ so far, our
explanation of the experimental puzzle becomes a prediction for
future $T_1$ measurements. In addition, the
spin-hot-spots picture suggests a way of altering $T_1$ via
band-structure modifications. Doping into the conduction band, for
example, may shift the Fermi surface away from some special symmetry
points and increase $T_1$. Similarly, alloys or
systems with reduced dimensionality will have spin
relaxation rates different from those of the corresponding elemental
or bulk metals, respectively. A general rule of thumb for
increasing $T_1$ of electrons in a conduction band is washing out the
spin hot spots from the Fermi surface.
In what follows we introduce the basic concepts of our theory
and discuss the results in a qualitative fashion.

If the periodic potential due to ions in a crystal lattice
contains spin-orbit
coupling (a term proportional to the scalar product of the orbital
and spin momentum operators, $\hat{{\bf L}}\cdot\hat{{\bf S}}$),
the electronic Bloch states are a mixture of spin up
$\mid\uparrow\rangle$ and down $\mid\downarrow\rangle$
species\cite{elliott54}:
\begin{eqnarray}
\Psi^{\uparrow}_{{\bf k}n}({\bf r})& = &[a_{{\bf
k}n}({\bf r})\mid\uparrow\rangle +b_{{\bf
k}n}({\bf r})\mid\downarrow
\rangle] \exp(i{\bf k}\cdot{\bf
r}),\\
\Psi^{\downarrow}_{{\bf k}n}({\bf r})& = &[a^*_{-{\bf k}n}({\bf
r})\mid\downarrow
\rangle -b^*_{-{\bf k}n}({\bf r})\mid\uparrow\rangle] \exp(i{\bf k}
\cdot{\bf r}).
\end{eqnarray}
The lattice momentum and band index are ${\bf k}$
and $n$, respectively, and $a_{{\bf k}n}({\bf r})$ and $b_{{\bf
k}n}({\bf r})$ are complex periodic functions with the period of the
lattice: if ${\bf G}$ denote the reciprocal lattice vectors, then
$a_{{\bf k}n}({\bf r})=\sum_{\bf G}a_{{\bf k}n}({\bf G})
\exp(i{\bf G}\cdot{\bf r})$ and
similarly for $b_{{\bf k}n}({\bf r})$. Both states have
the same energy $E_{{\bf k}n}$, as follows from time and space
inversion symmetry \cite{elliott54}; the numbering of bands is
therefore the same as without the spin notation.  The degenerate
states $\Psi^{\uparrow}_{{\bf k}n}$ and $\Psi^{\downarrow}_{{\bf
k}n}$ are chosen to represent electrons with spins polarized along
$z$ direction \cite{yafet63}:  $(\Psi^{\downarrow}_{{\bf
k}n}|\hat{S_z}| \Psi^{\downarrow}_{{\bf k}n})=
-(\Psi^{\uparrow}_{{\bf k}n}|\hat{S_z}| \Psi^{\uparrow}_{{\bf k}n})
< 0 $
and the off-diagonal matrix elements are zero. This condition implies
that $a_{{\bf k}n}({\bf r})$ have values close to one, while
$b_{{\bf k}n}({\bf r})$ are much smaller, decreasing with the decrease
of the strength of the spin-orbit interaction (with the exception of
the points where the spin-orbit interaction lifts a degeneracy).

Elliott\cite{elliott54} pointed out that ordinary
(spin conserving) impurity
or phonon scattering can induce transitions between
$\Psi^{\uparrow}_{{\bf k}n}$ and $\Psi^{\downarrow}_{{\bf k'}n'}$
(either their spin up or down amplitudes),
leading to the flip of a spin polarization and thus spin relaxation.
Simplifying assumptions lead to the formula\cite{prl}
\begin{eqnarray}\label{eq:3}
1/T_1\approx 4\langle b^2 \rangle/\tau,
\end{eqnarray}
where $\langle b^2 \rangle$ is the Fermi surface average of
\begin{eqnarray}\label{eq:4}
|b_{{\bf k}n}|^2=\sum_{{\bf G}}|b_{{\bf k}n}({\bf G})|^2,
\end{eqnarray}
and $\tau$ is the momentum relaxation time.
The weakness of the spin-orbit interaction makes the average
spin-mixing parameter $\langle b^2 \rangle$ much smaller
than one and $1/T_1\ll 1/\tau$. At low temperatures,
in a very pure sodium, for example,
$T_1$ can reach a microsecond\cite{kolbe71}, much
larger than momentum relaxation times which would reach a fraction of
a nanosecond in similar samples. Note that Eq. \ref{eq:3} implies a
series ``spin resistor'' model (scattering rates are additive)
for the spin relaxation, if $|b_{{\bf k}n}|^2/\tau$ is interpreted as
the spin-flip rate for the scattering from, or to state ${\bf k}n$. The
reason is that because $1/T_1\ll 1/\tau$, a single electron
experiences many different (``series'') scattering events before
changing its spin. Therefore the Fermi surface states with the highest
spin-flip rates count most. (A counterexample is conductivity
which is monopolized by the states with the lowest scattering rates.) 

In addition to the Elliott's mechanism, impurities and phonons affect the
spin relaxation in other ways. Impurities induce the spin-orbit
interaction which allows a direct transition between the spin up part
of $\Psi^{\uparrow}_{{\bf k}n}$ and the spin down part of
$\Psi^{\downarrow}_{{\bf k'}n'}$. The resulting spin relaxation
is independent of temperature (as is the Elliot's impurity
relaxation) and can be experimentally controlled.
Similar transitions can be induced by the phonon-modulated
spin-orbit interaction. This intrinsic effect leads, in principle,
to a spin relaxation that is as effective as the Elliott's
phonon-induced spin relaxation, but as we show below,
becomes unimportant in many polyvalent metals where the
Elliott's mechanism is enhanced by band structure.

In most cases the spin-orbit interaction can be incorporated into
the band structure as a perturbation, leading to the estimate
$\langle b^2 \rangle\approx (\lambda/\Delta)^2$. Parameter
$\lambda$ is some effective spin-orbit coupling and $\Delta$ 
is a typical (unperturbed) energy difference between neighboring
bands.  Monod and Beuneu \cite{monod79} estimated both $\lambda$
and $\Delta$ from their atomic values and found widely varying
results for different metals.  While this atomic substitution works
well for alkali and noble metals, other metals (Al, Pd, Mg, and Be)
seem to have $\langle b^2 \rangle$ much larger than calculated.  The
case of aluminum and sodium is instructive. These metals have similar
atomic numbers so their  atomic $\lambda/\Delta$ are also similar
(within 10\% \cite{monod79}). Yet their spin relaxation times differ
by two orders of 
magnitude\cite{prl,feher55,vescial64,schultz66,johnson85}!  The reason
is the very different band structure of the two metals 
that gives very different values for $\Delta$.  In sodium, $\Delta$ is 
of order $E_F$, the Fermi energy. In aluminum, $\Delta$ 
varies between zero (here perturbation theory for degenerate states 
gives $|b_{{\bf k}n}^2|\approx 0.5$) and $E_F$.  Remarkably, the 
regions with small $\Delta$ (the spin hot spots) have enough weight 
to significantly increase the average $\langle b^2\rangle$ beyond the 
naive estimate $(\lambda/E_F)^2$ which works so well for monovalent 
metals. A solid state environment greatly influences 
$\langle b^2 \rangle$ and $T_1$, leading to what we call the {\it 
band renormalization} of the spin-orbit mixing $\langle b^2 \rangle$.

Figure \ref{fig:1} shows our calculation \cite{prl} of the distribution
$\rho$ of the values attainable by $|b_{{\bf k}n}|^2$ over
the Fermi surface of aluminum.
The width of the distribution is impressive--almost seven decades!
Most states have $|b_{{\bf k}n}|^2$ below $10^{-5}$. Higher values
are much less frequent but they stretch up to about $10^{-1}$
(we lacked enough precision to reach the upper limit 0.5)
with $\rho$ decreasing linearly, as seen in the inset.  This long
tail, however, has a marked influence on the average $\langle b^2
\rangle \approx 2.0\times 10^{-5}$ which is an order of magnitude
larger than the value where $\rho$ is maximal.  The unusual character
of the distribution is further emphasised by the plot (also in
Fig.\ref{fig:1}) of $\rho$ for a hypothetical case of monovalent
aluminum.  The band structure parameters
(like the lattice structure or the form of the electron-ion
potential) are unchanged so that any difference in $\rho$ between
the two metals is caused solely by their different valencies, and
therefore different Fermi surface geometries.
The distribution of $b^2$ is now relatively
narrow with the average $\langle b^2 \rangle \approx 3.4\times
10^{-7}$ coinciding with the center of the distribution; this
value is about
fifty times smaller than the average for (trivalent) aluminum.
Adjusting for the density of states (monovalent aluminum would have
$1/\tau$ reduced $\approx 3^{1/3}$ times) the spin relaxation would
be about seventy times slower.  This is about the measured difference
between sodium, which has similar Fermi surface as monovalent
aluminum, and aluminum\cite{prl}.

The difference between the two very different forms of $\rho$ in Fig.
\ref{fig:1} is clearly attributed to the subtleties of the band
structure of aluminum. The
Fermi surface of monovalent aluminum
is nearly spherical and lies entirely within the first Brillouin zone;
the protuberances  near the zone boundaries do not touch the planes
as in the case of the noble metals. On the other hand, the Fermi
surface of aluminum crosses the Brillouin zone boundaries and
accidental degeneracy lines\cite{ashcroft63}. The distorted electronic
states near the crossings are responsible for the long
tail of $\rho$. Consider a band structure computed
without the spin-orbit interaction.  The spin-orbit interaction,
being a part of the periodic lattice potential can induce transitions
only between states whose ${\bf k}$ differ by a reciprocal lattice
vector, or, in a reduced-zone scheme, between the states with the
same ${\bf k}$ but different $n$. Let, for a state ${\bf k}n$ on the
Fermi surface, the closest band to $n$
is separated from $n$ by $\Delta$. The spin-orbit interaction mixes
the spin amplitudes from the two bands (more distant bands to $n$
will have smaller contribution and we neglect them) leading to the
spin-mixing parameter $|b_{{\bf k}n}|^2 \approx
(1-\Delta/\sqrt{\Delta^2+4\lambda^2})/2$.  Three cases can occur. (A)
For a general point on the Fermi surface, the band separation is of
order $E_F$, the Fermi energy, so that $\Delta\gg V_{SO}$ and
$|b_{{\bf k}n}|^2\approx (V_{SO}/E_F)^2$.
This is the case of monovalent aluminum: parameters $\lambda
\approx 3$ meV (same as for trivalent aluminum\cite{prl}) and $E_F
\approx 6$ meV give $|b_{{\bf k}n}|^2\approx 3\times 10^{-7}$
($\approx 10^{-6.6}$) in accordance with Fig. \ref{fig:1}.
In (trivalent) aluminum typical
spacings between bands are somewhat smaller than
$E_F\approx 12$ eV; the estimate of $\Delta\approx 3$ eV is in
reasonable agreement with Fig. \ref{fig:1}.
(B) If the state is close
to a Brillouin zone boundary associated with 
${\bf G}$, the band separation is
$\approx 2V_{\bf G}$ ($V_{\bf G}$ is the ${\bf G}${\it th} Fourier
coefficient of the non-spin part of the lattice potential). 
Since
typically $V_{\bf G}\gg V_{SO}$, $|b_{{\bf k}n}|^2\approx
(V_{SO}/2V_{\bf G})^2$; this can be a few orders larger than in (A).
Aluminum has $V_{111}\approx 0.2$ eV which gives $|b_{{\bf k}n}|^2$
about $6\times 10^{-5}$($\approx 10^{-4.2}$) coinciding with the
onset of the tail in Fig. \ref{fig:1}. Curiously, the Fermi  surfaces
of the noble metals too come into contact with some zone boundaries.
The noble metals, however, have unusually large $V_{\bf G}$
\cite{cohen70} so the estimate (A) works equally well.
Finally, (C) the spin-orbit interaction can lift the degeneracy of
two or more bands. The mixing of spins is complete and $|b_{{\bf
k}n}|^2\approx |a_{{\bf k}n}|^2\approx 0.5$. Even this case occurs in
aluminum, where the second and third band accidentally coincide at
some Fermi surface points (the accidental degeneracy points).
The neighborhood of these points is responsible for the long tail of $\rho$
in Fig.\ref{fig:1}. The spin hot spots are the states with the properties 
(B) and (C).

As Fig.\ref{fig:1} reveals, the spin hot spots essentially determine
the average $\langle b^2 \rangle$. We give a simple example
demonstrating that this is reasonable. Consider electrons in the
presence of a single Brillouin zone  plane that is
associated with the reciprocal vector ${\bf G}$. If the electron-ion
potential (or, rather, pseudopotential) is weak, the band structure
of such a system is found by using just two plane
waves, $\exp(i{\bf k}\cdot{\bf r})$ and $\exp[i({\bf k}-{\bf
G})\cdot{\bf r}]$; this is sometimes known as the two
orthogonalized-plane-wave (OPW) method\cite{harrison66}. The problem 
has axial symmetry along the direction of ${\bf G}$, so we conveniently 
shift
the origin of ${\bf k}$ to ${\bf G}/2$ and decompose ${\bf k}$ into
the components parallel (${\bf z}$) and perpendicular (${\bf r}$) to
${\bf G}$: ${\bf k}={\bf G}/2+{\bf r}+{\bf z}$. To further simplify
the notation, energy and momentum will have the units of
$\hbar^2(G/2)^2/(2m)$ and $G/2$, respectively; $m$ will be the electron 
mass. The band structure is formed by two energy levels with the dispersion
(see, for example, Ref. \cite{harrison66})
\begin{eqnarray}\label{eq:5}
E^{\pm}&=&1+r^2+z^2\pm\frac{1}{2}sgn(z)
\Delta (z), \\
\label{eq:6}\Delta (z)&=&\sqrt{16z^2+4V^2_{\bf G}}.
\end{eqnarray}
We allow $z$ which measures the distance from the plane, both
negative and positive values (repeated-zone scheme) and take
$E^{+}$ as the reference level. The Fermi
surface is then the revolution of the curve
\begin{eqnarray}\label{eq:7}
E^{+}(r,z)=E_F
\end{eqnarray}
about ${\bf G}$. If $z$ is negative (states inside the first
Brillouin zone) $E^{-}$ is the upper band, if $z$ is positive $E^{-}$
is smaller than $E^{+}$; the spacing between the two bands
$\Delta(z)$ depends only  on $z$.

The spin-orbit interaction will mix
the spin amplitudes of the two bands.  In most cases $\lambda \ll V_{\bf
G}$ and we can use the estimate (B):
\begin{eqnarray}\label{eq:8}
|b_{{\bf k}n}|^2\equiv |b_{z}|^2 \approx \left 
[\lambda/\Delta(z)\right ]^2.
\end{eqnarray}
A proper evaluation of $|b_{{\bf k}n}|^2$ would, instead
of effective $\lambda$, have the matrix elements of the
spin-orbit interaction between the band states with
energies $E^-$ and $E^+$. Such matrix elements, in general, 
depend on both $r$ and $z$,
and for spin quantized not along ${\bf G}$, also on the angle of
revolution $\phi$. In particular, the spin mixing would vanish
at ${\bf k}={\bf G}/2$
(that is, $r=z=0$). This follows from
group-theoretical arguments\cite{jaro2} or, more simply, from the
fact that the spin-mixing matrix elements between the plane waves
$\exp(i{\bf k}\cdot{\bf r})$ and $\exp[i({\bf k}-{\bf
G})\cdot{\bf r}]$, out of which the band states are formed, are
proportional to the vector product ${\bf k}\times({\bf k}-{\bf G})$;
this product vanishes for ${\bf k}={\bf G}/2$.
Incidently, this vanishing of spin mixing 
is another reason why the band structure does not 
affect $T_1$ for the noble metals despite the fact that some of their 
states come into contact with zone boundaries at ${\bf G}/2$.
Introducing effective $\lambda$ therefore overestimates the region
around ${\bf G}/2$. In most interesting cases of polyvalent
metals, however,  the Fermi surface crosses the plane far enough from
this point (that is, if $z=0$ radius $r$ is a significant
fraction of $G/2$) and our approach is justified.

The Fermi surface average of $|b_z|^2$ given by  Eq. \ref{eq:8} is
\begin{eqnarray}\label{eq:9}
\langle b^2 \rangle = \frac{1}{z_2-z_1}
\int_{z_1}^{z_2}dz \frac{\lambda^2}{\Delta^2(z)}.
\end{eqnarray}
The limits of the integration, $z_1$ and $z_2$, are the minimum and
maximum value of $z$ reached by the Fermi surface; they are 
obtained by putting $r=0$ and solving Eq. \ref{eq:7} for $z$. We
are interested in how $\langle b^2 \rangle$ depends on the Fermi
surface geometry.  The Fermi energy which determines
$z_1$ and $z_2$ will now be a variable parameter (simulating
doping).  Although the integral in Eq. \ref{eq:9} can be evaluated
analytically it is more instructive to consider some limiting cases only.
The numerical estimates will be done with the band-structure
parameters appropriate for aluminum so that the results can be directly compared
with the full band structure calculation of
Fig.\ref{fig:1}. The parameters are  
${\lambda} \approx 4.3\times 10^{-4}$ and ${V}_{\bf
G}= {V}_{111}\approx 0.035$ (we now use the dimensionless
units for energy). Although the aluminum Fermi surface
crosses both $(111)$ and $(200)$ zone boundaries, we choose as
$V_{\bf G}$ the Fourier
coefficient $V_{111}$ which is much smaller than $V_{200}$ and has
therefore greater impact on $\langle b^2 \rangle$.

If $E_F$ is small, the Fermi surface is a small,
almost undistorted sphere.  Both $z_1$
and $z_2$ are close to $-1$ so we can neglect $V_{\bf G}$ in Eq.
\ref{eq:6} and obtain $\langle b^2 \rangle \approx
\lambda^2/16\approx 1.2\times 10^{-8}$.  
This number is not affected by the presence of the Brillouin zone
boundaries and can be thought of as an atomic limit for the
spin mixing. When $E_F$ is large enough
for the Fermi surface to cross the zone boundary, $z_2\agt-V_{\bf
G}$ and the small region of the states around the plane gives
the dominant contribution to $\langle b^2 \rangle$. Indeed, when
taking the integral in Eq. \ref{eq:9} 
from $-V_{\bf G}$ to $V_{\bf G}$ only,
$\Delta(z)\approx 2V_{\bf G}$ and $\langle b^2 \rangle \approx
\lambda^2/(4V_{\bf G}k_F)$; the Fermi momentum 
$k_F=\sqrt{E_F}$ is given by $z_2-z_1\approx 2k_F$. 
Before substituting numerical values we note that there are eight 
$(111)$ planes in aluminum. They contribute independently to $\langle 
b^2 \rangle$ giving the estimate $\langle b^2 \rangle \approx 
8\times 10^{-6}$.  The full 
calculation in Fig. \ref{fig:1} gives the value which is more than twice 
larger. This is explained partly by our neglect of the $(200)$ planes 
and, more important, by ignoring the accidental 
degeneracy points.

Figure \ref{fig:2} plots the full dependence of 
$\langle b^2 \rangle$ on $E_F$ within the two OPW approximation
for the eight (111) aluminum planes. 
Doping, or increasing the size of the Fermi surface can increase
$\langle b^2 \rangle$ by almost three orders of magnitude! As the
Fermi surface expands towards the zone boundaries, the Bloch states that
are closest to the planes (they have smallest $\Delta$) become more
and more important. Once the Fermi surface reaches the plane (at
the energy $1-V_{\bf G}\approx 0.97$) the average $\langle b^2 \rangle$
goes through a maximum and then slightly decreases. The decrease
is tied to the saturation of the integral in Eq. \ref{eq:9}:
only the states in the small neighborhood of the plane are dominant and 
$\langle b^2 \rangle\sim 1/(z_2-z_1)$. The quantity  $z_2-z_1$ is
up to a numerical factor the density of states. 
As Fig. \ref{fig:1} shows, 
in this toy calculation
monovalent aluminum would have $\langle b^2 \rangle$ about fifteen
times smaller than aluminum, in good agreement with the results of
Fig. \ref{fig:1}.

The qualitative understanding of the above estimates for the impact
of a Brillouin zone boundary on $\langle b^2 \rangle$ is quite simple.
We need to estimate the probability with which an electron in its
random walk on the Fermi surface jumps into a state close to a
Brillouin zone boundary. Alternatively, we can ask what is the
portion of the free-electron Fermi surface states that have $\Delta
\alt 2V_{\bf G}$ (these states will be significantly perturbed by
the presence of the boundary).  The answer is $\sim V_{\bf G}$ and 
the reason that this number is linear in $V_{\bf G}$ (the linearity
is crucial) is that $\Delta$ increases
linearly away from the plane, as in Eq. \ref{eq:6}. The average value 
$\langle b^2 \rangle$ is then $\sim V_{\bf G}\times \lambda^2/V^2_{\bf G} 
\approx \lambda^2/V_{\bf G}$. The enhancement of the spin mixing
due to the presence of a single Brillouin zone boundary is then 
$\sim 1/V_{\bf G}$, as in Fig. \ref{fig:2}.

To estimate how accidental degeneracy
points affect $\langle b^2 \rangle$ is more difficult\cite{silsbee83}. 
Consider two
bands coming into contact at a single point $R$, different from the symmetry
points, on the Fermi surface.  As noted in Ref. \cite{harrison66}
the band spacing $\Delta$ grows linearly as we go away from $R$ in almost
all directions (the exception is the direction along the corresponding 
accidental degeneracy line). We need to divide the region around $R$
in two: the first has states with $0\alt \Delta \alt 2\lambda$, the second
with $2\lambda\alt \Delta \alt 2V_{\bf G}$. Since $\Delta\sim \delta k$,
where $\delta k$ measures the distance from $R$, there are $\sim \lambda^2$
points in the first region, all with (approximately) the same value
of $b_{{\bf k}n}\sim 1$ from estimate (C). The average $\langle b^2 \rangle$
therefore scales as $\lambda ^2$. The second region is different in that
we have $b_{{\bf k}n}$ depending on $\Delta$ [estimate (B)]. The average
$\langle b^2 \rangle$ is thus proportional to 
\begin{eqnarray}
\langle b^2 \rangle \approx \int_{\lambda}^{V_{\bf G}}
d\Delta \Delta \frac{\lambda^2}{\Delta^2}.
\end{eqnarray}
This integral evaluates to $\lambda^2\ln(V_{\bf G}/\lambda)$ giving a weak 
enhancement of about $\ln(V_{\bf G}/\lambda)\approx 4$. More accurate
evaluation based on a four OPW approximation would give an additional
enhancement of $(V_{200}/V_{111})^2\approx10$ \cite{prl}. As $V_{\bf G}$ 
becomes smaller the importance of accidental degeneracy points diminishes.

Another type of a spin hot spot, not relevant for aluminum though,
is the region around a special symmetry point. It may happen
that the Fermi energy coincides with a set of two or more degenerate
levels at a symmetry point. This is the case, for example,
  of the fcc palladium
and platinum, whose Fermi surfaces go through the fcc L points 
\cite{papa86}. If the
spin-orbit interaction lifts this degeneracy, the renormalization
of $\langle b^2 \rangle$ can be significant. This effect is, however,
not easy to estimate qualitatively. For the fcc W point we find that
the enhancement of $\langle b^2 \rangle$ relative to the atomic value
of $\sim \lambda ^2$ is about $V_{\bf G}/\lambda$\cite{jaro2}.
This can range from a ten to a thousand.

Our final note concerns the hexagonal Mg and
Be, where the deviation of $T_1$ from the simple estimates is most
striking \cite{monod79}. We argue that this is also a manifestation
of the band renormalization of $\langle b^2\rangle$. Without the spin-orbit
interaction, all the states at the hexagonal faces of the first
Brillouin zone 
of a simple hexagonal structure are degenerate \cite{ashcroft76}.
The spin-orbit interaction lifts this degeneracy \cite{elliott54} (except
at some symmetry points and lines),
presumably by the amount $V_{\bf G}\lambda$, the
largest second-order term containing the spin-orbit interaction
(any first order term vanishes since the structure factor
associated with the hexagonal faces is zero\cite{ashcroft76}).
The contribution to $\langle b^2 \rangle$ of the points where
the Fermi surface intersects the hexagonal faces is
$\sim V_{\bf G}\lambda$:
the characteristic value $|b_{{\bf k}n}|^2
\sim 1$, times the area of the affected part of the Fermi surface,
$V_{\bf G}\lambda$. The enhancement measured in terms
of $\lambda^2$ is then $V_{\bf G}/\lambda$; this can be as large as
a thousand for light elements like Mg and Be.

We thank P. B. Allen and M. Johnson for helpful discussions.
This work was supported by the U.S. ONR.

\newpage

\begin{figure}
\caption{Calculated distribution $\rho$ (in arbitrary units)
of the spin-mixing parameters
$|b_{{\bf k}n}|^2$ for aluminum. The corresponding average
$\langle b^2\rangle\approx2.0\times10^{-5}$ is indicated by a solid arrow.
The linear tail of the distribution is shown in the inset.
The dashed line shows what the distribution would
be if aluminum were monovalent ($\langle b^2\rangle\approx3.4\times10^{-7}$,
dashed arrow).
}
\label{fig:1}
\end{figure}

\begin{figure}
\caption{Two OPW calculation (with the parameters $V_G$ and $\lambda$
suitable for aluminum) of the average spin-mixing parameter
$\langle b^2 \rangle$ as a function of the Fermi energy $E_F$. The values
of $\langle b^2 \rangle$ that correspond to monovalent (open circle,
$E_F=0.81$) and trivalent (filled circle, $E_F=1.7$) aluminum
are $4.1\times 10^{-7}$ and $6.2\times 10^{-6}$, respectively. The
vertical line indicates the case where the free-electron 
Fermi sphere comes into contact with the Brillouin zone boundaries.
}
\label{fig:2}
\end{figure}


\begin{references}
\bibitem{johnson93a} M. Johnson, Science {\bf 260}, 320 (1993);
J. Magn. Magn. Mater. {\bf 140-144}, 21 (1995); {\it Ibid.} {\bf 156}, 321 (1996);
 Mater. Sci. Eng.  B {\bf 31}, 199 (1995).
\bibitem{prinz95} G. Prinz, Phys. Today {\bf 48}, No. 4, 58 (1995).
\bibitem{pines55} The longitudinal ``spin-lattice'' relaxation time
$T_1$ equals, at least for cubic metals, the transverse ``dephasing''
relaxation time $T_2$, as shown in D. Pines and C. P. Slichter,
Phys. Rev. {\bf 100},
1014 (1955).
\bibitem{divincenzo95} D. P. DiVincenzo, Science {\bf 270}, 255 (1995).
\bibitem{kikkawa97} J. M. Kikkawa, I. P. Smorchkova, N. Samarth,
and D. D. Awschalom, Science {\bf 277}, 1284 (1997); J. M. Kikkawa
and D. D. Awschalom, Phys. Rev. Lett. {\bf 80}, 4313 (1998).
\bibitem{prl} J. Fabian and S. Das Sarma, Phys. Rev. Lett. {\bf 81}, 5624 (1998). 
\bibitem{monod79} P. Monod and F. Beuneu, Phys. Rev. B {\bf 19},
911 (1979); F. Beuneu and P. Monod, Phys. Rev. B {\bf 18},
2422 (1978).
\bibitem{elliott54} R. J. Elliott, Phys. Rev. {\bf 96},
266 (1954).
\bibitem{yafet63} Y. Yafet, in {\it Solid State Physics}, edited by F.
Seitz and D. Turnbull (Academic, New York, 1963), Vol. 14.
\bibitem{silsbee83} R. H. Silsbee and F. Beuneu, Phys. Rev. B {\bf 27},
  2682, (1983).
\bibitem{kolbe71}W. Kolbe, Phys. Rev. B {\bf 3},
320 (1971).
\bibitem{feher55} G. Feher and A. F. Kip,
Phys. Rev. {\bf 98}, 337 (1955).
\bibitem{vescial64} F. Vescial, N. S. Vander Ven, and R. T. Schumacher,
Phys. Rev. {\bf 134}, A1286 (1964).
\bibitem{schultz66} S. Schultz, G. Dunifer, and C. Latham, Phys. Lett.
{\bf 23}, 192 (1966); D. Lubzens and S. Schultz, Phys. Rev. Lett. {\bf 36},
1104 (1976).
\bibitem{johnson85} M. Johnson and R. H. Silsbee, Phys. Rev. Lett.
{\bf 55}, 1790 (1985); Phys. Rev. B {\bf 37}, 5326 (1988).
\bibitem{ashcroft63} N. W. Ashcroft, Phil. Mag. {\bf 8}, 2055 (1963).
\bibitem{cohen70} M. L. Cohen and V. Heine, in {\it Solid State Physics},
edited by H. Ehrenreich, F. Seitz, and
D. Turnbull (Academic, New York, 1970), Vol. 24, p. 183.
\bibitem{harrison66} W. A. Harrison, {\it Pseudopotentials in the Theory
of Metals}, (Benjamin, New York, 1966).
\bibitem{jaro2} J. Fabian and S. Das Sarma (unpublished).
\bibitem{papa86} D. A. Papaconstantopoulos, {\it Handbook of the Band 
Structure of Elemental Solids}, (Plenum, New York, 1986).
\bibitem{ashcroft76} N. W. Ashcroft and N. D. Mermin, {\it Solid
State Physics}, (Saunders, New York, 1976).
\end{references}
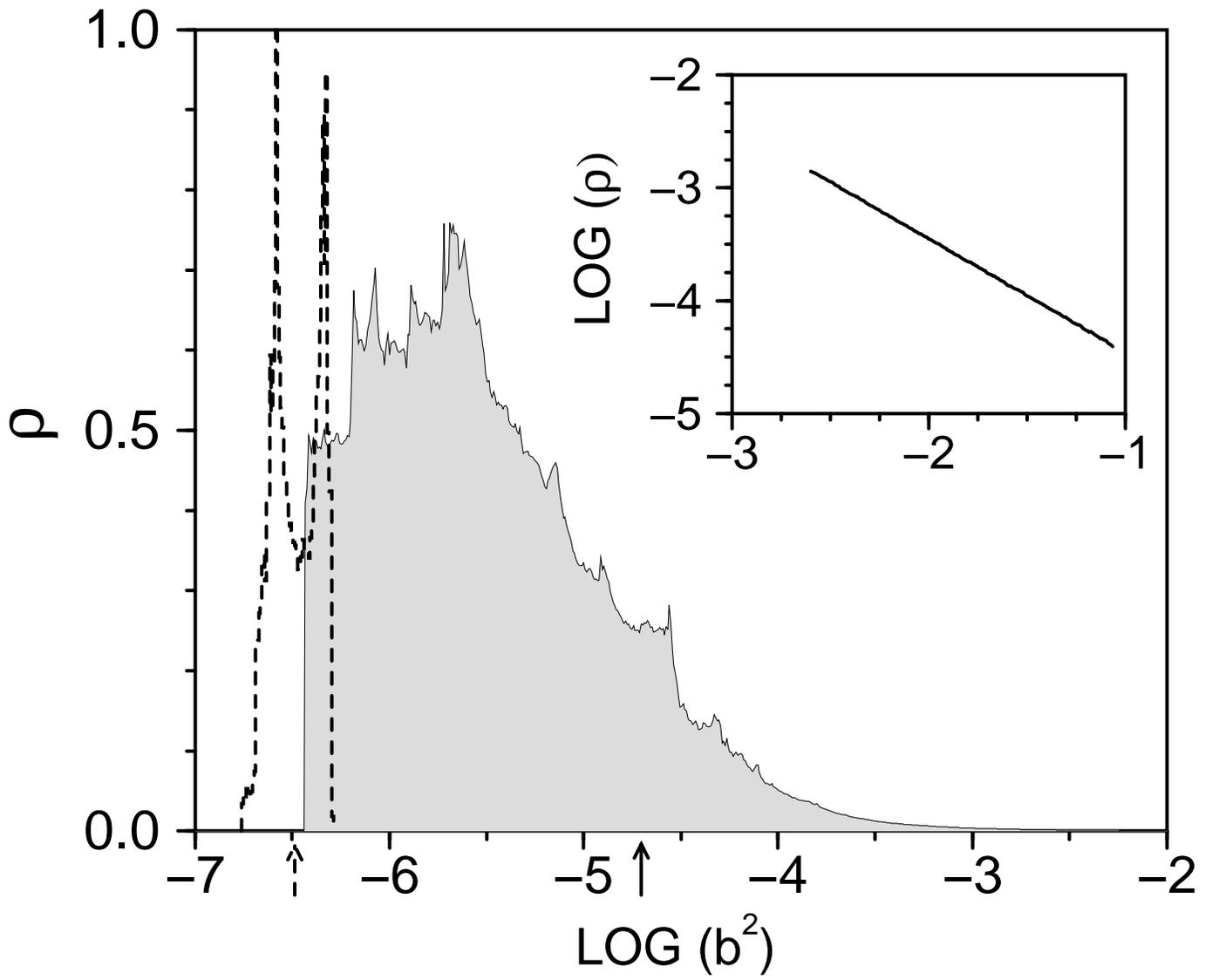
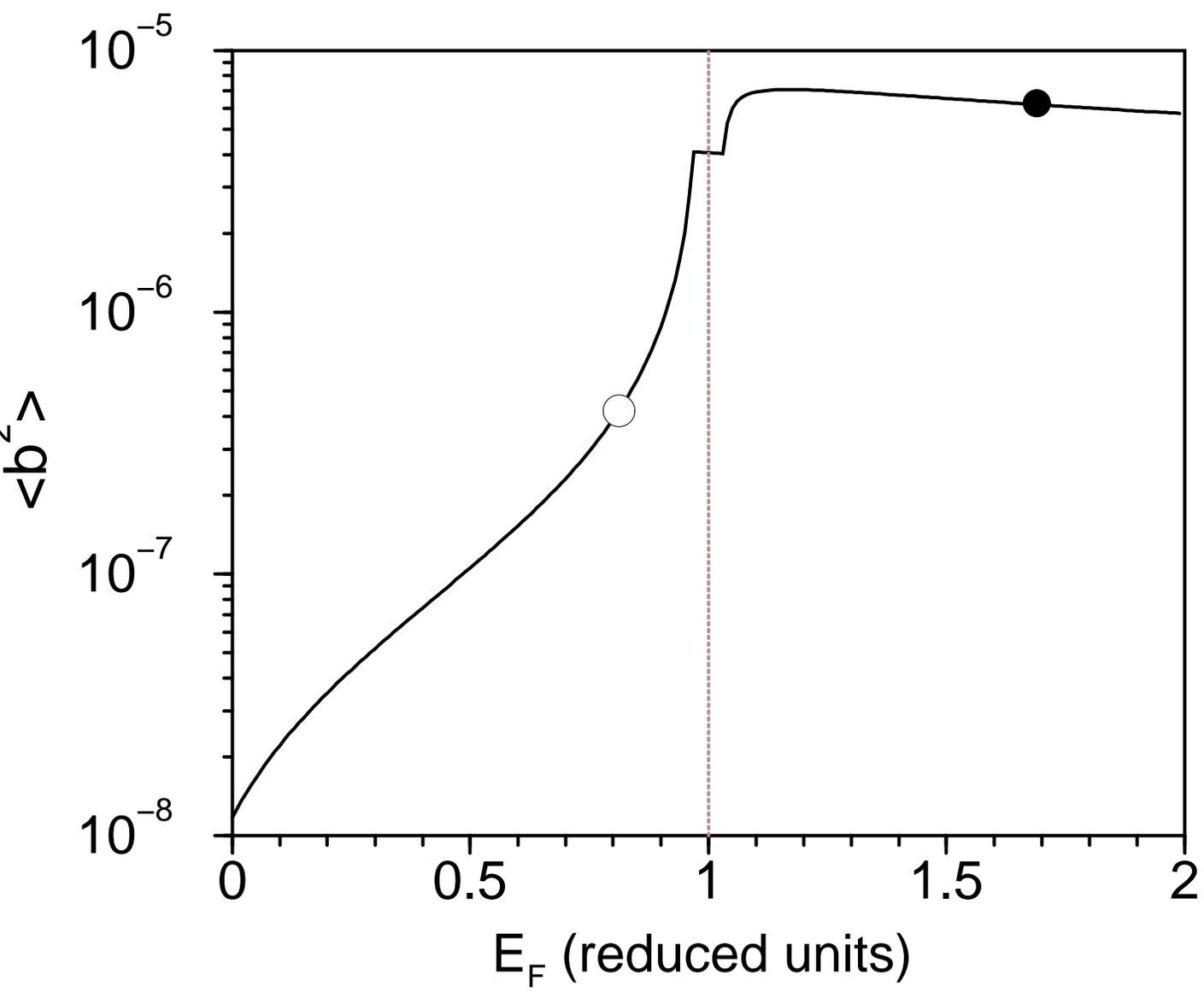
\end{document}